# Preparation Methods and Applications of Biomimetic Membranes

Ajit Seth[a], Sajal K. Ghosh[a] and Veerendra K. Sharma[b, c*]

[a] *Department of Physics, Shiv Nadar Institution of Eminence, NH 91, Tehsil Dadri, G. B. Nadar, Uttar Pradesh, India, 201314*
[b] *Solid State Physics Division, Bhabha Atomic Research Centre, Mumbai 400085, India*
[c] *Homi Bhabha National Institute, Mumbai 400094, India*
**Email: sharmavk@barc.gov.in; vksphy@gmail.com*

## Abstract

Model biomembrane systems play a crucial role in advancing biomedical research by providing simplified yet effective platforms for exploring complex biological mechanisms. These systems span a wide range of scales, from single-molecule-thick lipid monolayers to micron-sized giant unilamellar vesicles. Their efficacy and applicability largely depend on selecting an optimal model and an appropriate synthesis process. This chapter offers a comprehensive description of conventional synthesis techniques, highlighting their limitations across various model membrane systems. Additionally, it provides an overview of biophysical studies on biomimetic membranes and explores key biological applications, including drug delivery, membrane-protein interactions, and biosensing.

## 1. Introduction

All six kingdoms of life, namely animalia, plantae, fungi, protista, archaea, and bacteria, owe their existence to the unceasing functioning of biological membranes. These membranes enclose a cell and are selectively permeable to foreign entities entering the cell environment. They comprise a double leaflet of amphiphilic phospholipids or glycolipids molecules with embedded proteins and sterols.

In the previous century, the structure and functioning of these cellular membranes have been explained through multiple models. The first breakthrough came with the Dutch scientists Gorter and Grendel discovering the existence of a lipid bilayer in the cellular membrane [1]. They discovered that the molecular area occupied by the extracted lipids from red blood cells on a Langmuir trough was double the cellular area measured using microscopy. Danielli and Davson, further proposed the existence of a protein layer throughout the membrane, with the moieties attached to the polar head group of lipids [2] . The confirmation of dispersed protein moieties throughout the membrane led to the proposal of Fluid Mosaic model by Singer and Nicolson [3]. According to this model, the protein and sterol entities extend throughout the double leaflet of lipid molecules across a homogenous fluid state, with varying symmetry of particular leaflets in the distribution of lipids. Further, in 1997, Simons et al. [4] and Brown et al. [5] contradicted the existence of a homogenous fluid lipid phase as the Fluid Mosaic model proposed . They showed that the phase-separated microdomains, called lipid rafts with an explicit composition, existed throughout the membrane with strikingly distinct dynamics from the adjoining lipids in the fluid phase. This model with phase-separated lipid rafts was widely accepted and is still directly impacting the understanding of the complex functioning of the membrane.

The biological membranes are responsible for numerous vital metabolic, signalling, transport, and structural functioning of the cell. This includes the regulated transportation of substances through barrier formation between the cytosol and extracellular environment aided by diffusion and osmosis. These membranes maintain the cell equilibrium by controlling the ion balance, pH, and hydration levels, providing the necessary stability for the cell. The protein receptors present in these membranes identify and transmit the electrical and chemical signals generated by hormones and neurotransmitters [6]. This assures effortless communication between the cell and other signalling molecules. Further, the cell membrane maintains sufficient binding between adjacent cells and the extracellular matrix to form cohesive tissues successively. Moreover, these membranes can facilitate protein-protein interactions by modifying their thickness and lateral organization. However, exploring these multifaceted characteristics of real biological membranes through in vivo or in vitro measurements or even theoretical approaches requires the consideration of their immense complexity. Thus, these approaches should include considering a delicate balance of a few hundred components in a biological system in an aqueous medium while simultaneously ensuring the non-damaging nature of the probing techniques. Therefore, technically obtaining a detailed and error-free analysis of the biological membrane is severely challenging. Hence, researchers explore the closest possible biological scenario using multiple biomimetic approaches. This involves exposing distinct membrane components to multiple entities in a controlled environment.

Modern biomimetic membranes have proven to be highly efficient in reconstructing the diverse functionalities of biological membranes. This includes exploring their selective permeability towards proteins, ions, and other nanoparticles. These systems allow the probing of membranes at multiple length scales with high spatial and sequential resolutions, which leads to discovering the underlying mechanisms behind multiple biological processes. Further, the applications of biomimetic membrane systems are incredibly diversified, including drug delivery, effortlessly selective biosensors, scaffolds for tissue growth, etc. Moreover, the selectively permeable characteristics of these membranes find multiple industrial applications, including antimicrobial channels [7] and even optoelectronic devices [8] . In this regard, this chapter provides a comprehensive summary of various synthesis techniques and the applicability of common biomimetic membrane systems. The following section explores widely used preparation methods, highlighting their respective advantages and limitations. The subsequent sections delve into advanced biophysical studies and key applications, offering a deeper understanding of the potential and impact of these systems.

## 2. Common Methods to Prepare Biomimetic Membrane Systems

The biomimetic membrane systems constitute a diverse set of models, including lipid monolayers, lipid bilayers constituting supported lipid bilayers (SLBs) and tethered lipid bilayers (TLBs), liposomes constituting unilamellar vesicles (ULVs) and multilamellar vesicles (MLVs), and lipid nanodiscs. Figure 1 shows the arrangement of lipid molecules in the aforementioned multiple biomimetic membrane systems. The following section further discusses the multiple preparation techniques of these biomimetic membrane systems.

## 2.1 Lipid Monolayers

Monomolecular thick layer of lipid molecules constitutes an excellent system for quantifying lipid molecules' thermodynamical, electrical, optical, and structural properties when exposed to protein, drug, or other relevant molecules [9]. These lipid molecules are amphiphilic in nature as they are composed of a hydrophilic head and a hydrophobic hydrocarbon tail. This essential hydrophobicity of the nonpolar tails allows them to get uniformly dissolved in nonpolar solvents like chloroform. These nonpolar solvents with dissolved lipid molecules lead to their rapid spreading at the air-water interface, with the tails suspended in air while the lipid heads are immersed in water. These molecules tend to spread in a gas-like arrangement to minimize their free energy. The evaporation of the nonpolar solvents eventually leads to the formation of an immiscible one molecular thick layer at the air-aqueous interface known as the Langmuir monolayer (Figure 1(a)). This tendency of lipid molecules to instinctively form Langmuir monolayers provides a deep insight into various molecular phenomena that otherwise would have been unattainable through other biomimetic models, as phospholipid monolayers allow precise control over the composition and type of lipid molecules being used. These lipid monolayers can be transferred onto different substrates to quantify the morphological changes induced to the lipid packing.

Primarily, in a Langmuir trough (Figure 2(a)), the lipid molecules are compressed using barriers, and the fall in the surface tension of the air-aqueous interface by the dispersed lipid molecules is quantified as the increase in surface pressure. This increase in surface pressure is plotted against the area available to each lipid molecule at the interface and is termed the surface pressure-area isotherm. The surface pressure-area isotherm also quantifies the phase changes the lipid molecules undergo upon compression to eventually form a continuous monolayer at the air-water interface (Figure 2(b)). Here, the term phase represents the varying aggregation of molecules and the corresponding intermolecular interactions being governed by a combination of van der Waals forces among the hydrocarbon tails and the coulombic forces between the lipid head groups. For instance, when the molecules are initially dispersed and rapidly spread at the air-aqueous interface, the lipid molecules are in random orientations with no sufficient adhesion to develop any measurable surface pressure. In this phase, the molecules follow a gas-like behaviour, and hence the region of the isotherm with a gradually reducing area per molecule and no increase in the pressure represents the gaseous phase. Upon further compression, the molecules tend to come closer, forming a liquid expanded phase (LE) (Figure 2(c)), which is reflected in the increase in surface pressure. In this LE phase, the molecules tend to have some level of adhesion, but the hydrocarbon tails are still in their random orientations. After a further reduction in the area per molecule, a plateau region is observed in the isotherm. This plateau region signifies another phase transition of the molecules, and this new phase is termed the Liquid Condensed (LC) phase. In this phase, the continuous strong adhesion among the lipid molecules leads to a sharp increase in the surface pressure, which signifies a compact confirmation of the hydrocarbon tails. However, this unique tail confirmation is strictly dependent on the size, shape, and polarity of the lipid molecules. Further compression of the monolayer leads to a solid-like arrangement of molecules, suggesting their tightly packed arrangements. Finally, an additional compression triggers the molecules to override and form bilayers or multilayer structures, with some molecules being lost into the subphase, causing a collapse of the monolayer.

The successful collection of surface-pressure area isotherm curves also allows for extracting multiple distinct parameters comprehending the biophysical changes occurring at the air-aqueous interface [10]. In addition to quantifying the area per molecule for the different phases of the monolayer, the slope of a surface pressure-area isotherm provides information about the compression modulus of these monolayers. This variation of the compression modulus explicitly explains the packing and ordering of molecules at the air-aqueous interface, eventually discovering the changes induced to the elasticity of the monolayer while switching from one monolayer to another and even from one component to a multi-component system. Moreover, in a multiple-component system, the area under the curve of a surface pressure-area isotherm can be utilized to evaluate the excess Gibbs energy from mixing these molecules. These excess Gibbs energy values successfully quantify the deviation of the mixed systems from an ideal behaviour and the corresponding magnitude of interaction between the components and their thermodynamic stability [11].

While phospholipid monolayers provide an option to carry out an array of measurements for the self-assembly of multiple entities with cellular components, they have their own set of limitations. It is a highly simplified biomimetic model compared to tremendously complex biological cell membranes that typically exist as lipid bilayers. Moreover, monolayer studies are based on sensitive measurements as they exceedingly require highly specific environmental conditions.

### 2.2 Supported Lipid Bilayers (SLBs)

Supported lipid bilayers (SLBs) are another popular biomimetic membrane systems that involve the immobilization of lipid bilayers on extremely clean and hydrophilic solid substrates (Figure 1(b)). The existence of a solid substrate allows for the exploitation of the lipid bilayer systems through multiple surface-sensitive measurements, comprehending the topographical features of membranes, including atomic force microscopy (AFM), quartz crystal microbalance (QCM) measurements, and surface scattering measurements [12]. These topographical explorations lead to a deeper understanding of multiple membrane functionalities, including adhesion characteristics of a membrane, lipid lateral diffusions, and even lipid-protein binding affinities [13]. Furthermore, the additional stability provided by a solid substrate to the lipid bilayer makes them tremendously valuable for the development of advanced platforms for biosensing applications. Also, SLBs exist in an aqueous environment that allows these model systems to closely mimic the complex aqueous environment of real biological cell membranes.

The synthesis of SLBs is generally done through vesicle fusion process, the Langmuir-Blodgett/Schaeffer (LB/LS) deposition, or a combination of the Langmuir-Blodgett and vesicle fusion processes. The vesicle fusion process to form SLBs involves the adsorption of small unilamellar vesicles onto a solid substrate driven by a delicate balance of the adhesion, repulsion, and hydration forces existing between the vesicle and the substrate [14]. Figure 3(a) shows the formation of an SLB through the vesicle fusion process. In this process, the desired substrate is initially placed in a closed chamber filled with an aqueous solution, followed by the addition of a vesicle suspension. Primary, separate vesicles adhere to the substrate, and the interactions between the vesicles and the substrate directly control the adhesion phenomenon. If these interactions are strong enough, the adhesion process stresses out the vesicles, causing them to deform and rupture onto the substrate, forming an isolated bilayer island. Meanwhile, the other vesicles can get ruptured and get laterally attached to the end of this bilayer. These interactions between the substrate and the vesicles are occasionally inadequate to initiate the rupturing process. Then, the concentration of vesicles in the solution is increased gradually to induce the required stresses among adjacent vesicles and cause rupturing. Upon reaching a critical value of this concentration, the crowding of multiple vesicles adhered to the substrate leads to the rupturing of vesicles due to the enhanced osmotic and membrane stresses [14] . Eventually, a continuous lipid bilayer is formed with some trapped fluid between the bilayer and the substrate at random locations. To achieve a clean sample, one has to wash out the excess vesicles from around the supported bilayer by exchanging the solution in the chamber, keeping the bilayer continuously hydrated.

Although the vesicle fusion process has long been used to form SLBs successfully, a few inevitable drawbacks are associated with this method. The issue arises with synthesizing stable vesicles with desired compositions to obtain the targeted SLBs. After obtaining the desired vesicles, the substrates and the aqueous environment conditions must be highly favourable for the vesicles to approach and adhere to the substrates. Moreover, the formation of SLBs is directly dependent on the spontaneous rupturing capabilities of the vesicles upon coming in contact with the substrate. This adhesion and the corresponding rupturing of the vesicles are significantly affected by the hydrophilicity and surface roughness of the substrate, which in turn leads to inhomogeneity in the obtained SLBs [15]. The formation of homogenous SLBs also depends on the rupturing and adhesion of multiple adjacent vesicles on the substrate, which could induce pore formation and nonuniformity in the obtained bilayer.

As mentioned earlier, the Langmuir Blodgett/Schaefer (LB/LS) technique can be utilized to obtain SLBs. This method involves dip-coating a lipid monolayer on a hydrophilic substrate on the upstroke cycle, followed by sufficient drying and then subsequent deposition of another lipid monolayer to eventually form an SLB [16]. The nomenclature of the technique depends on the deposition of the second monolayer. If the second monolayer is deposited similarly to the first monolayer by vertical dipping of the substrate into the subphase, the technique is termed LB/LB deposition (Figure 3(b)(i)). If the second monolayer is deposited after the lateral inversion of the substrate to uptake all the lipid molecules across the substrate area at once (Figure 3(b)(ii)), the technique is called LB/LS deposition. Furthermore, SLBs can also be formed by combining the LB and vesicle fusion method (Figure 3(b) (iii)). In this process, the first monolayer is dip-coated following the LB deposition, and the second monolayer is deposited through vesicle fusion. Primarily, the monolayer-coated substrate is allowed to incubate in the unilamellar vesicle-containing solution. The fusion of a vesicle on this monolayer strictly depends on the charge and composition of the lipid molecules. When the vesicle comes in contact with the monolayer, the intense hydrophobic interactions between the lipid tails and the vesicle induce multiple defects in the vesicle. The formation of these defects causes the vesicle to break down and the subsequent release of the monolayer [17]. The deposition of individual monolayers in these approaches assures excellent control over the composition of lipid molecules being attached to the substrate. This ultimately provides the option to obtain asymmetric lipid bilayers, representing a close mimic of a native cellular membrane, as this asymmetricity is widespread in the physiological cellular bilayers.

One key advantage of this approach is the stable surface and smooth morphology, which allow atomic force microscopy to capture topographical features with atomic-scale resolution. This enables the visualization of nanometer-scale domains and defects characteristic of biological membranes. Nevertheless, the direct interaction between the bilayer's lower leaflet and the solid substrate can introduce artifacts [18]. These include suppressing the main phase transition, increasing lipid diffusion at short distances, and inhibiting bilayer undulations, which may affect the system's biological relevance. Additionally, this form of a biomimetic model cannot incorporate the rheological measurements achievable for other model membrane systems.

While vesicle fusion and Langmuir Blodgett/Schaefer have been regularly utilized for the synthesis of SLBs, a relatively novel, simpler, and versatile approach known as the solvent-assisted lipid bilayer (SALB) method has been recently established [19]. The SALB method overcomes limitations associated with traditional SLB fabrication techniques, such as vesicle fusion, by using lipid solutions in water-miscible organic solvents. This method involves minimal sample preparation and the usage of basic microfluidics to obtain diverse SLBs on a wide array of solid supports such as silicon oxide, gold, graphene, and even titanium oxide [20]. Initially, a lipid mixture of a desired concentration is first dissolved in a water-miscible organic solvent such as isopropanol. Parallelly, the substrate is cleaned and then rendered hydrophilic via oxygen plasma or UV/ozone treatment. A microfluidic chamber is assembled with the treated substrate, and the system is pre-filled with aqueous buffer to remove air bubbles. The chamber is then flushed with pure isopropanol, after which the lipid solution is introduced and optionally incubated to allow lipid adsorption. The organic solvent is subsequently exchanged with aqueous buffer at a controlled flow rate, prompting self-assembly of the lipids into a bilayer. The chamber is finally rinsed with additional buffer to remove excess lipids, forming a uniform, stable SLB.  However, like other synthesis processes, this method also suffers from its specific drawbacks. It is highly sensitive to the choice of organic solvent and lipid composition, with isopropanol yielding the best results, while other solvents may lead to defective or incomplete bilayers. Also, the solvent exchange rate has to be extremely precise for the formation of homogenous SLBs, as too rapid or too slow exchange rate can critically induce the development of SLBs containing defects.

### 2.3 Tethered Lipid Bilayers (TLBs)

While SLBs constitute a lipid bilayer affixed directly onto a solid substrate, a tethered lipid bilayer (TLB) consists of a tethering unit connecting the membrane to the solid substrate (Figure 1(c)). This tethering unit generally comprises polymers acting as soft support, providing cushioning to the bilayer and extraordinary stability to this model membrane system [21]. Adding a soft tethering offers greater flexibility to the bilayer as it reduces the substrate-membrane interactions and allows for exploring membrane dynamics, making TLB a superior biosensing platform compared to SLB. Also, the usage of TLBs permits the incorporation of transmembrane proteins, as the proteins do not come in contact with the substrate, simplifying protein-membrane interaction studies.

The synthesis process for TLBs involves functionalizing the substrate with a polymer with excellent control over its thickness and composition [22]. On top of these polymers, the bilayer can be deposited through two different mechanisms, the first being the existence of an independent polymer cushioning layer below the bilayer (Figure 4(a)), and the second case includes polymer molecules coupled with the inner leaflet of the bilayer on one end and attached to the substrate on the other end (Figure 4(b)), acting as the tethered support [22]. In the first kind of TLB, the successful deposition of a lipid bilayer onto the independent polymer-cushioned substrate depends strictly on the uniformity and elasticity of the chosen polymer. Additionally, the polymer should be inert enough not to hinder the self-assembly process of the lipid molecules. Once the desired cushioning of the polymer has been obtained, the lipid bilayer can be easily deposited following the conventional vesicle fusion or the LB/LS deposition method discussed earlier. In the polymer-lipid coupled TLBs, the lipid molecules generally mixed with lipopolymers are deposited onto a substrate using the LB technique [23]. However, the capability of these lipopolymer entities to get adsorb onto the substrate directly affects the stability of the final TLBs. To overcome this, a highly reactive linker layer is deposited onto the substrate before attaching the lipopolymer monolayer. This linker layer facilitates the stable attachment of the lipopolymer molecules, eventually forming a covalently bonded TLB. Although the first monolayer must be necessarily deposited through the LB method, the second monolayer can be deposited through the LB, LS or vesicle fusion technique, as discussed earlier.

While TLBs provide an additional fluidity to the lipid bilayer, their synthesis process is highly complex, consisting of many sensitive steps. The successful incorporation of the lipid membranes onto the polymer surface requires an exclusive selection of hydrophilic polymers. Furthermore, the water-induced swelling behavior of polymers can cause defects in the bilayer structure, leading to the loss of uniformity and integrity of the membranes.

## 2.4 Liposomes

Liposomes constitute another versatile model membrane system set that contributes to basic research and clinical applications. These are spherically shaped vesicles entrapping an aqueous solution bounded by either a single or multiple concentric lipid bilayer. Based on these characteristics, these liposomes are classified into two categories, the first being the multilamellar vesicles (Figure 1(d)) and the second being the unilamellar vesicles (ULVs) (Figure 1(e)).

The easiest way to prepare liposomes is the thin-film hydration method. In this method, lipid molecules are dispersed in a nonpolar solvent in a glass flask, and the consequent drying of this solution is used to obtain dried lipid films on the flask. These lipid films are dried planar lipid bilayer aggregates, and these bilayers tend to be in their lowest free energy states [24]. These dried lipid aggregates tend to swell when exposed to the hydration process due to the hydrophobic repulsions between the hydrocarbon tails and water. Therefore, upon receiving some external energy in the form of mild agitations or stirring, these swelled lipid films overcome the free energy barrier to detach away and rapidly form multilamellar vesicles (MLVs) following the famous Bangham method (Figure 5) [25]. However, these rapidly obtained MLVs lack homogeneity in terms of liposome size and lamellarity. Hence, these MLVs are subjected to multiple freeze/thaw cycles to induce agitation, forming homogenous MLVs with large encapsulation capacities [26]. The unique lamellarity in MLVs ensures a controlled release of the encapsulated drug molecules, reducing their side effects [27] . In addition to that, several aqueous sections and hydrophobic bilayers enable the MLVs to encapsulate and carry both hydrophilic and hydrophobic molecules [28] . These properties and the rapid formation of MLVs ensure their repeated usage in the dermatology and the cosmetic industry to mimic skin lipid compositions [29]. However, the existence of multiple compartments and the corresponding heterogeneity in MLVs can hinder their applicability for exploring the permeability and fusion of bilayers [30]. These issues arise because of the opposing ion flux among multiple bilayers, making it complicated to discover the interactions between different bilayers during the membrane fusion process.

The successful formation of MLVs paves the way for the formation of unilamellar vesicles (ULVs), which constitute another significant set of model membrane systems. These lipid assemblies comprise a single lipid bilayer enclosing an aqueous core within them. ULVs are generally classified based on their sizes where small unilamellar vesicles (SUVs) are of the order of ~20-100 nm, large unilamellar vesicles (LUVs) are around ~100-1000 nm, and all the vesicles having sizes greater than ~1000 nm constitute giant unilamellar vesicles (GUVs) [31]. ULVs are exceptionally stable, and their synthesis allows exclusive control over their sizes. This leads to their regular exploitation for various biophysical studies and multiple biological applications, including gene, drug, and even deoxyribonucleic acid (DNA) delivery purposes. The membrane-active proteins can effortlessly interact with these liposomes due to their existence in an aqueous environment. This facilitates the exploration of multiple underlying mechanisms behind the membrane-protein interactions [32].

While MLVs are the liposomes in their lowest energy state, they require significant energy from external sources to overcome the free energy barrier to eventually form ULVs [24]. Depending on the application, the synthesis techniques of the ULVs depend directly on the required size. For instance, the most common methods for obtaining ULVs are sonication, extrusion, and electroformation. In the sonication process, high-pressure waves are generated using acoustic energy inside solutions containing MLVs to cause their breakdown. This process can either follow the bath sonication process, where the vessel containing the lipid solution is dipped in a water/ice bath, or the probe sonication process, where a metallic probe is immersed in the lipid solution to deliver high-energy waves [33]. The sonication time, temperature, and the corresponding intensity of the applied energy decide the size of the obtained vesicles [33]. Furthermore, the sonication process is solvent-free and highly time-efficient as it causes the rapid disintegration of MLVs. The sonication process is one of the widely used processes to obtain ULVs; however, the high energy waves lead to the formation of ULVs with varying sizes and can even cause lipid degradation, leading to the loss of functionality of the obtained liposomes [34] . Furthermore, in the case of probe sonication, the lipid suspensions can be contaminated by the metal fragments released by the probe immersed in the suspension.

Extrusion is another conventional process to obtain ULVs from large-sized liposomes [31]. It involves the manually repeated passage of the lipid suspension solution through polycarbonate membranes with predefined pore sizes to obtain ULVs. This process provides a uniform size of ULVs with excellent reproducibility. The ULVs are formed due to the shearing forces induced by the polycarbonate membrane at very high pressures [35] . Although extrusion is known for its high reproducibility, but the process is relatively slow, generally manual, and challenging to carry out at high pressures. The repeated passage of the lipid suspension through the porous polycarbonate membrane can lead to some loss of materials, limiting its usage for large-scale sample preparations. Moreover, reducing the vesicle size at high temperatures during the extrusion process for desired applications is challenging.

While sonication and extrusion have been extensively used for the formation of ULVs, typically SUVs and LUVs, electroformation is another method that is exclusively used for the formation of GUVs [36]. The first step in this process involves spreading lipid solutions constituting an organic solvent on the surface of conducting indium tin oxide-coated glass electrodes. After the evaporation of the used organic solvent, the dried lipid film is then hydrated in an aqueous environment. The final step involves the application of an alternating electric field to the electrodes, causing the lipid film to swell and form GUVs. Compared to other methods for GUV synthesis, electroformation is a rapid and time-efficient process. Furthermore, the size and composition of the target vesicles can easily be controlled by tuning the applied electric field and the sample environment. While electroformation has been highly efficient for synthesizing GUVs, it still suffers from some setbacks. This technique necessitates optimal control of the applied electric energy as high electric field values can disrupt the lipid bilayers, causing disintegrations [36]. Also, minor changes in the multiple protocols involving lipid drying and hydration can cause the GUVs to lose homogeneity, and the practice of using expensive electrodes limits their scalability.

Liposomes, either in the form of MLVs or ULVs, offer one of the possible mimics to the physiological cell and are readily utilized for multiple applications ranging from drug delivery to the food industry. However, like other biomimetic systems, they have their own set of limitations. For instance, MLVs, due to their lamellarity, offer extremely low encapsulation volume for load delivery purposes and have limited long-term stability due to their tendencies to get aggregated under varying physiological conditions. Further, the processes involving the synthesis of ULVs are complex and restrict their usage for large-scale applications. Additionally, the obtained ULVs are specific and homogenous, failing to mimic the signature heterogeneity of real biological systems.

### 2.5 Nanodiscs

While lipid monolayers, bilayers, and liposomes together constitute the traditional set of biomimetic systems, nanodiscs represent a relatively modern category. Nanodiscs constitute disc-shaped phospholipid bilayers enclosed by belt-like structures. Traditionally, these structures were composed entirely of amphipathic proteins known as membrane scaffold proteins (MSPs); however, multiple studies have also reported the usage of peptides and polymers [37]. Nanodisc model membrane systems are efficient for isolation, purification, and even structural and functional characterization of membrane proteins. They also provide control over the stability of the oligomeric states of the reconstituted membrane protein [38] .

Generally, the exploration of a membrane protein's characteristics requires a stable suspension of detergent-protein-lipid micelles. However, the presence of detergents in these suspensions hinders multiple assay techniques and structural characterizations. The synthesis process of nanodiscs simply involves isolating the desired membrane protein in a detergent micelle, followed by adding chosen lipids and membrane scaffold protein. The mixed detergent molecules are then removed through dialysis or other adsorption methods leading to the spontaneous formation of nanodiscs with the membrane scaffold protein (MSP) simultaneously assembling into the discoidal lipid bilayer [39] . Figure 6 shows the formation of a traditional nanodisc enveloping a membrane protein bacteriorhodopsin (bR) extracted from a halobacterium's purple membrane [40]. In addition to providing an excellent platform for lipid-protein interactions, these nanodiscs are outstanding drug carriers due to their flat discoidal structure, which provides excess loading area and improved half-life compared to conventional liposomes. Nanodiscs also allow the loading of hydrophobic and hydrophilic molecules, including drugs, ribonucleic acids (RNA) in the hydrophobic bilayer core, and peptides through surface-bound interactions [41] .

Although nanodiscs have significant advantages, like other biomimetic systems, they have their own limitations. Selecting detergent molecules to form stabilized nanodiscs requires vigorous exploration as one detergent may induce protein denaturation for one specific MSP that might be suitable for another. Researchers have explored replacing these detergent entities with styrene-based polymers. However, these styrene-based polymers with strong absorbance in the UV region and low stability have been observed to hinder multiple spectroscopy-based studies of nanodiscs.

The above-mentioned preparation methods for different biomimetic systems allow for the isolation and exploration of diverse biological problems under well-defined conditions. This is achieved by performing different biophysical measurements on these biomimetic systems, which further allow the exploration of diverse interaction mechanisms behind multiple biological processes ranging from the molecular to the macroscopic level. These biophysical approaches, ranging from scattering measurements, thermodynamic measurements, electrical measurements, spectroscopy, and microscopy to force measurements and calorimetry, provide detailed insights into the physical properties and interactions of biological components, which are often difficult to observe directly in vivo.

## 3. Biophysical Studies of Biomimetic Membrane Systems

In the previous section, we discussed conventional synthesis methods and their drawbacks for different biomimetic systems. Here, a few examples of biophysical studies performed using these biomimetic systems are discussed, which eventually help comprehend the underlying mechanisms behind multiple physiological processes.

In a lipid monolayer system, any uptake of the protein or drug molecules and even nanoparticles mixed in the aqueous medium to the lipid monolayer is directly translated onto the measured isotherm. The Langmuir Blodgett setup allows for an articulate measurement of the surface pressure-area isotherm, which quantifies the thermodynamic properties of a lipid monolayer. A parallel quantification of the electrical properties of the monolayer is possible through the measurement of the surface potential of the lipid molecules across the interface. This surface potential-area isotherm measurement works on the assumption of considering every lipid molecule as an individual dipole with a specific dipole moment [42]. With the compression process, fluctuations are induced in the orientation and packing of these lipid molecules. This varies the dipole density in a defined area, and the surface potential is measured following the Helmholtz model, which assumes a parallel plate capacitor across the interface [43]. The measurement utilizes a vibrating capacitor model where a vibrating plate is positioned just above the air-aqueous interface, and a counter electrode is submerged below in the subphase to detect the potential differences. Moreover, the flawless dispersibility and the self-assembly of lipid molecules to form monolayers at the air-aqueous interface facilitates multiple structural characterizations, including X-ray and neutron scatterings. These scattering techniques allow the extraction of detailed information about the lipid molecules' structural organization, including their out-of-plane orientations and the in-plane lattice arrangements. In addition to scattering techniques, fluorescence and Brewster angle microscopy (BAM) measurements can be performed on the lipid monolayers. These measurements subsequently visualize the mesoscopic morphology, phase behaviour, and structure of lipid domains when the monolayer achieves the liquid condensed phase.

A study has recently compared the self-assembly of graphene-based nanomaterials, such as graphene oxide (GO) and reduced graphene oxide (rGO), around phospholipid monolayers with distinct electrostatic characteristics [44]. It effectively utilizes the Langmuir Blodgett trough to quantify the thermodynamical, electrical, and structural changes GO and rGO induced to the zwitterionic (dipalmitoyl-sn-glycero-3-phosphocholine, (DPPC)) and cationic (1,2-distearoyl-sn-glycero-3-ethylphosphocoline, (DSEPC)) phospholipid monolayers. Surface pressure-area isotherms, surface potential-area isotherms, and in-situ X-ray scattering measurements were performed at the air-aqueous interface using a liquid diffractometer. The surface pressure-area isotherm (Figure 7(a)) concluded an increased affinity of GO towards the cationic lipid monolayer than rGO, as confirmed by the increased compaction and higher compression modulus values obtained for the DSEPC/GO monolayer than the DSPEC/rGO monolayer. Meanwhile, rGO induced enhanced compaction in the zwitterionic DPPC monolayer compared to the same monolayer dispersed over the GO dispersed subphase. The surface potential values (Figure 7(b)) that were simultaneously recorded supported the conclusions drawn from the surface pressure measurements. It was observed that the GO nanoflakes, constituting a higher negative charge than rGO due to the excess oxygenated species on their surface, tend to cause an additional reduction in the surface potential values for the cationic phospholipid monolayer than the zwitterionic monolayer. Additionally, the X-ray reflectivity (XRR) (Figure 7(c)) and grazing incident X-ray diffraction (GIXD) measurements (Figure 7(d)) performed on these monolayers confirmed the assembly of GO near the hydrophilic head region of cationic lipid molecules and the penetration of rGO nanoflakes into the hydrophobic tail region of zwitterionic lipid monolayer. It was concluded that the interaction of GO nanoflakes with the lipid monolayers is readily governed by electrostatics, whereas for rGO nanoflakes, hydrophobic interactions dominate their interplay with lipid molecules.

In a recent study, in 2023, Villanueva and group investigated the interfacial behavior and enzymatic activity of a purified recombinant chitinase, Chi18-5, against chitosan-DPPC mixed Langmuir monolayers to mimic organized substrate membranes [45]. The structural and catalytic properties of this enzyme were evaluated at different pH levels (5 and 7), revealing that pH significantly affects its conformation, surface activity, and catalytic efficiency. Chi18-5 showed higher catalytic activity at pH 5, adopting a more flexible structure conducive to aggregation and enhanced interfacial interaction with chitosan-DPPC monolayer. The surface pressure changes allowed for the exploration of thermodynamic changes that the enzyme produces over the organized substrate film. The corresponding surface potential measurements parallelly confirmed the role of interfacial adsorption and the conformational rearrangements of the enzyme molecules at the interface in varying the enzyme activity at one pH corresponding to another. Eventually, BAM and AFM measurements confirmed the significant role of pH in modulating the aggregation behaviour of Chi18-5 at the interface.

In 2008, Ghosh et al. investigated the role of a plasma membrane lipid phosphatidylinositol-4,5-bisphosphate ($PIP_2$) in the vesicle fusion process through X-ray scattering studies and further quantified the interaction of synaptic vesicles with the SLBs (Figure 8(a)) composed of dioleoyl-snglycero-3-phosphatidylcoline (DOPC) and $PIP_2$ [46] . Small unilamellar DOPC and DOPC/PIP2 vesicles were prepared and eventually fused onto cleaned (100)-silicon wafers to form the SLBs. The interaction of synaptic vesicles, purified from rat brains, with the DOPC SLBs upon introduction of $PIP_2$ was quantified through X-ray reflectivity measurements (Figure 8(b)) and the fitted data, which is shown in Figure 8(c). It was eventually concluded that the mixing of $PIP_2$ into the DOPC SLB led to an enhanced attachment of synaptic vesicles onto the SLBs, as confirmed by the consistent increase in the increase in the electron densities of the head and tail region of the lipid molecules (Figure 8(d) and 3(e)).

In 2018, Kurniawan et al. successfully formed asymmetric lipid bilayers on mica surfaces, comparing the LB/LB and LB/LS deposition processes [16] . The first monolayer constituted entirely of 1,2-dipalmitoyl-sn-glycero-3-phosphoethanolamine (DPPE) molecules dip-coated through the LB deposition. The asymmetricity is introduced into the bilayer by depositing the outer monolayer constituting DOPC–DPPC–cholesterol (1:1:1) molecules once following the LB deposition and then again following the LS deposition. This comparative study utilized the AFM measurements to conclude the higher smoothness and uniformity of the LB/LB deposited SLB over its LB/LS counterpart. While the LB/LS technique can provide control over the lipid compositions, one major drawback of this process is the requirement of a particular Langmuir Blodgett apparatus. Moreover, only the gel phase SLBs deposited through this method tend to have considerable stability for long-term measurements.

TLBs with an independent polymer layer have been discussed by Bhattacharya et al. where they synthesized TLBs of 1,2-dipalmitoyl-sn-glycero-3-phosphocholine (DPPC) by spin-coating poly(acrylic acid) (PAA) on 3-aminopropyl-triethoxysilane (APTES) functionalized silicon wafers [47]. This PAA layer acted as the soft cushioning unit for the LB/LS deposition of the DPPC bilayer on silicon wafers (Figure 9(a)). In addition to that, X-ray reflectivity measurements (Figure 9(c)) were performed on the TLBs to comprehend the underlying mechanism behind the ionic liquid-induced bacterial cell death. The ionic liquid molecules were found to increase the electron density (Figure 9(d)) and reduce the thickness (Figure 9(e)) of the bilayer in both gel and fluid phases, confirming the perturbed assembly of the biomimetic cellular membrane.

The second kind of TLB with polymer-coupled monolayer has been discussed in another study by Lin et al. [23]. The synthesis process involved the usage of a lipopolymer 1,2-distearoyl-sn-glycero-3-phosphoethanolamine-N-[folate(polyethylene glycol)-5000] (DSPE-PEG5000) and the phospholipid 1-palmitoyl-2-oleoyl-sn-glycero-3-phosphocholine (POPC). The LB/LS technique was utilized to form the two different kinds of TLBs, a polymer-tethered lipid bilayer with varying changes of the lipopolymer concentrations and another membrane system with distinct regions of extremely low and high lipopolymer concentration in the same TLB. The first leaflet of the bilayer was composed of POPC with fixed mixing ratios of DSPE-PEG5000, and the second monolayer was composed entirely of POPC. Fluorescence microscopy and AFM were deployed to image and observe the topography of the synthesized TLBs. It was observed that the TLBs had lipopolymer concentration-specific buckling structures. These structures were formed due to lateral stress between the lipopolymer and phospholipid molecules. Moreover, the shape, lipid diffusivity, and elasticity of these structures were found to be dependent on the lipopolymer concentrations.

In a work, Kaushik et al. utilized the gel and fluid phase MLVs of a brain sphingolipid called N-octadecanoyl-d-erythro-sphingosylphosphorylcholine (BSM), which is abundantly found in the myelin sheaths wrapping the axons of neurons [48] . In the study, the small-angle X-ray scattering (SAXS) measurements were performed to comprehend the structural fluctuations caused by AMT molecules on the MLVs of BSM lipids in the gel and fluid phases (Figure 10). The data collected were fitted based on the work of Pabst et al. to obtain various structural parameters as detailed elsewhere [49] . In the gel phase, they found that the bilayer thickness decreased for the MLVs with AMT in the aqueous solution, which was attributed to the increase in the surface area of the molecules without showing significant variations in the volume of the hydrophobic core, inducing a change in the intrinsic curvature (Figure 10(a)). Further, the estimated electron density curves suggested that the lipid in the gel phase (high hydrophobicity) allows a higher number of AMT molecules to partition and perturb the chain region (significant change in electron density of tail region) as hydrophobic interaction is the primary driving force of their interaction (Figure 10(b)). However, the flexible and randomized lipid tails in the fluid phase could accommodate fewer AMT molecules with less disruption, resulting in a lesser reduction in the chain length (Figure 10(c) and 10(d)). Further, they also concluded that the absorption of the AMT molecules induces additional flexibility in the lipid membrane.

A temperature-dependent study by Bota et al. was performed on the ultrasonication process of DPPC lipid suspensions to synthesize ULVs [50]. They utilized time-resolved synchrotron SAXS experiments to quantify the phase changes of the lipid lamellae. The corresponding morphology of the suspension was visualized through Transmission electron microscopy combined with freeze-fracture (FF-TEM). For the lipid gel phase (20°C), ultrasonication induced high-temperature regions leading to irregular bilayer fragments (LBFs) formation. However, uniform ULVs were obtained for the lipid membrane in the ripple and liquid crystalline phases.

The changes in the size and morphology of dimethyldioctadecylammonium bromide or chloride (DODAX, X = Br or Cl) vesicles were quantified by Rusli et al., upon mixing with three different monovalent salts, namely sodium chloride (NaCl), sodium bromide (NaBr) and lithium chloride (LiCl) [51]. The salt-added DODOAX vesicles were prepared in two different ways, one involving the addition of salts pre-extrusion to the lipid suspension and another where the salts were added after the extrusion process. Here, the polydispersity and size of the obtained ULVs were measured through dynamic light scattering (DLS) and the morphology was observed through Cryo-Transmission electron microscopy (Cryo-TEM). For the pre-extrusion salt addition condition, DODOAX was observed to maintain quasi-spherical ULVs, while for the post-addition case, the final morphology of the vesicles was dominated by a mixture of undesired morphologies.

Sharma and coworkers, in 2015, explored the role of cholesterol and the lipid phase in modulating the interaction of an antimicrobial peptide melittin with ULVs of DMPC [52]. To observe the membrane dynamics, quasielastic neutron scattering (QENS) measurements were performed on the ULVs of DMPC in both gel and fluid phases, with and without the addition of melittin. The data revealed that adding melittin to DMPC ULVs significantly affected the membrane dynamics. In the gel phase, melittin induced additional flexibility and increased the lateral motion of the lipids, which acted as a plasticizer. It is generally observed due to the formation of transmembrane pores. In the fluid phase, the interaction of melittin with head groups of lipids leads to the stiffening of the membrane, restricting the lipid lateral motion. However, for cholesterol-supplemented vesicles, the destabilizing effects of melittin on the membrane dynamics disappear entirely, confirming the vital role of cholesterol in protecting and stabilizing the membrane.

Another study by Sharma et al. in 2019 discussed the effect of aspirin, ibuprofen, and indomethacin, three commercial nonsteroidal anti-inflammatory drugs, on the ULVs of DMPC in both the gel and fluid phase [53]. They utilized a combination of SANS and neutron spin echo (NSE) techniques to quantify the modulations induced by the drug molecules in the structure and dynamics of the DMPC lipid bilayer. SANS measurement concluded that individually incorporating all three drug molecules into the ULVs decreased the bilayer thickness. Similarly, NSE data showed that the bilayer's bending rigidity and compressibility modulus decreased significantly, irrespective of the drug molecule used. However, the degree of reduction in these parameters was observed to be directly dependent on the lipid phase and type of the drug molecule used. In the fluid phase, aspirin caused the maximum reduction in the bilayer thickness, bending rigidity, and compressibility modulus, while indomethacin caused the least. While aspirin reduced the bilayer thickness the most in the gel phase, similar to the fluid phase, ibuprofen showed the maximum effect on bending rigidity and compressibility modulus in this phase.

Zhang et al. in 2024 improved upon the conventional electroformation method (Figure 11(a)) and synthesized DPPC GUVs and even multilayered giant vesicles in physiological saline [54] . They modified the standard electroformation device by adding an insulating layer of Teflon in the middle of the chamber, dividing it into two independent parts (Figure 11(b)). This was done to prevent the free movement of ions between the chambers, which traditionally hindered the swelling of lipid membranes. Further, they showed that the improved device could synthesize GUVs even in buffer and ionic solutions (Figure 11(d)), which wasn't possible through the conventional method (Figure 11(c)), leading to the formation of small vesicles and lipid clumps in physiological saline. Furthermore, these giant unilamellar and multilayered vesicles successfully encapsulated plasmid DNA inside them (Figure 11(e)) and closely mimic the eukaryotic cells as confirmed by the fluorescence and laser confocal microscopy analysis.

In 2021, Johansen and coworkers synthesized five different nanodiscs made with dimyristoylphosphatidylcholine (DMPC) and different MSPs varying in size, charge, and circularization (Figure 12(a)) using the cholate-assisted method [55] . This method involves drying lipid film in a glass vial and adding a buffer solution containing sodium cholate to the dried lipid film [56] . This is followed by adding MSP to the cholate-solubilized phospholipid in the desired ratio. Further, this solution is incubated with amberlite XAD-2, a hydrophobic copolymer of styrene-divinylbenzene resin, to extract the cholate molecules, leading to the spontaneous formation of nanodiscs. These nanodiscs were then characterized by a combination of SAXS, differential scanning calorimetry (DSC), and time-resolved small-angle neutron scattering (TR-SANS). Primarily, temperature-controlled (10°C to 37°C) SAXS measurements were performed for all the five nanodiscs. It was observed that the gradual increase in temperature caused the first minima of the SAXS curve to shift consistently to lower q-values for all the measurements, confirming the reorganization of lipid molecules from the tightly packed gel phase to a more fluid-like arrangement. The SAXS data for one conventional nanodisc and one supercharged nanodisc is shown in Figure 12(b). Further, the DSC measurements performed for all the different nanodiscs indicated that lipid packing remains unaffected even when the charge and curvature of the nanodiscs are changed (Figure 12(c)). This was confirmed by the computation of the linear thermal expansion coefficient values for all the bilayers, which turned out to be strikingly comparable, irrespective of the size and supercharging of the nanodiscs. Moreover, comparing the nanodisc DSC data with pristine DMPC LUVs confirmed the perturbed packing of lipids near the region closest to MSP while the central lipids were more tightly packed than the pristine LUVs. While the novel supercharged nanodiscs exhibit similar lipid packing as other nanodiscs, TR-SANS showed that this supercharging reduced the lipid exchange rates within the nanodiscs (Figure 12(d)), which could lead to some desorption of lipids and require additional exploration.

## 4. Applications of Model Biomembrane Systems

With their simplified and precise approach, model membrane systems have enabled researchers to explore multiple physiological processes outside of the biological milieu. These model systems find multiple applications ranging from single molecular studies to in vivo delivery of drugs and even chemotherapy. The following section discusses some of the applications of the model biomembrane systems.

### 4.1 Drug Delivery

Liposomes and nanodiscs have been readily used for drug delivery purposes, including hydrophilic and hydrophobic drug molecules. The existence of an aqueous core in liposomes, including ULVs and MLVs, and the lipid bilayer assembly in nanodiscs allows for the encapsulation and loading of drug molecules. For instance, in a study by Xiong et al., sterically stabilized liposomes were loaded with multiple anti-cancerous drug molecules, which had an improved therapeutic effect for melanoma tumors [57] . In addition, liposomes have been observed to protect the encapsulated drug molecules from degradation, providing excellent stability. A work by Niu and coworkers showed that liposomes containing sodium glycocholate (SGC), sodium taurocholate (STC), or sodium deoxycholate (SDC) loaded with insulin were exceptionally able to slow down and sustain the delivery action for diabetic rats even for a period of over 20 hours [58] . Moreover, liposomes allow for evaluating the interaction mechanism and even the delivery of other membrane-active molecules, including proteins, peptides, enzymes, antibiotics, genes, and nucleic acids. In this regard, a study by Shaban et al. showed liposome's targeted microRNA delivery mechanism in treating glioblastoma, one of the most prevalent brain cancers in adults [59] . These liposomes encapsulated microRNAs and were also able to increase the sensing capabilities of the cells to ionize radiations, effectively improving therapeutic efficiency. Furthermore, liposomes have also been observed to encapsulate protein and peptide molecules specifically for oral delivery purposes. This safeguards the convenience and cost-effectiveness of the oral drug administration process with maximum patient compliance.

Apart from liposomes, lipid nanodiscs have outstanding biocompatibility and drug-loading capacity. They have also principally contributed to drug delivery, including peptide, gene, and vaccine delivery. To this end, Kuai and colleagues showed the development of high-density lipoprotein–mimicking nanodiscs loaded with a chemotherapeutic agent doxorubicin (DOX) [60] . These loaded nanodiscs were able to trigger cancer cell death while simultaneously maintaining non-toxicity towards the surrounding environment. In another study, Chen and coworkers fabricated two distinct hybrid nanodiscs with cyclic Arginine-Glycine-Aspartic acid (cRGD) peptide-functionalized on the nanodisc's edge in one case and the top planar surface in another [61]. These nanodisc were then loaded with small interfering RNA and compared for their gene delivery efficacies. Additionally, Najafabadi et al. discussed the vaccination delivery proficiency of synthetic high-density lipoprotein nanodiscs against high aldehyde dehydrogenase, a biomarker for cancer stem cells [62] . These nanodisc, when coordinated with immunotherapy, had extremely efficient antitumor tendencies and were even able to prolong the animal survival probabilities.

### 4.2 Membrane Protein Research

This section discusses the contribution of a few studies utilizing SLBs, including TLBs, ULVs, and nanodiscs, towards advancing membrane protein research. The biochemical and structural analysis discovers the crucial interactions between membrane proteins and lipids that define their biological behaviour. Solid supports and large surface area in SLBs and TLBs allow for multiple characterization of membrane proteins, including surface plasmon resonance, quartz crystal microbalance measurements, X-ray and neutron scatterings, fluorescence microscopy, and atomic force microscopy. For example, Pace and the group successfully conducted a hybrid study by fusing native lipid vesicles containing transmembrane proteins with synthetic vesicles to eventually form polymer-supported TLBs on glass surfaces [63] . They were able to comprehend the structure and orientation of β-secretase 1, a transmembrane protein, in the hybrid SLBs using total internal reflection fluorescence imaging. In 2021, Luchini et al. discussed the degradation effect of spike protein extracted from SARS-COV-2 on SLBs of lipid mixtures containing natural and synthetic lipids [64] . Using neutron reflectometry, they showed that the spike protein molecules, when dispersed in the solution with SLBs, could destructively erode the lipid molecules from SLBs. In another study, Banerjee and the group explored the cholesterol-controlled aggregation of Amyloid β, a protein responsible for the development of Alzheimer's disease, on mixed POPC SLBs [65]. They concluded that the presence of cholesterol in SLBs drastically increases the aggregation rate of Amyloid β, and these aggregates could also dissociate from the bilayer into the solution. These results confirm the importance of lipid composition in modulating the membrane-protein interactions by controlling the aggregation and orientation of protein entities. Furthermore, Qian and group quantified the interaction of an anionic lipid 1,2-dimyristoyl-sn-glycero-3-phosphoglycerol (DMPG) and Amyloid β Peptide [66]. While SANS measurements showed that the incorporation of Amyloid β into the ULVs of DMPG is strictly limited to the interfaces without any penetration into the lipid tail region, the membrane dynamics were explored through Quasielastic Neutron Scattering (QENS) measurements. QENS results concluded that the gradual addition of Amyloid β could only modify the lateral motion of lipid molecules but not the fast internal motions. However, these altercations observed in the membrane dynamics could affect multiple membrane functionalities, including disruptions in cell signalling and energy transduction pathways.

Apart from SLBs, the stability of nanodisc also provides a platform for exploring membrane-protein interactions. These disc-shaped lipid bilayers effortlessly encapsulate MSPs, which allows the formation of a smooth suspension of proteins coupled with lipids. For instance, Bocquet utilized the surface plasmon resonance (SPR) technique to perform a real-time monitoring of a thermostabilized and truncated version of the human adenosine (A2A) G-protein-coupled receptor (GPCR) being inserted in a lipid bilayer nanodisc [67] . They showed that for small molecule binding studies using SPR, nanodiscs are exceptional in providing considerable stability to membrane proteins, closely mimicking the physiological environment. Kynde and group extracted structural information about a membrane protein coupled with a nanodisc using SAXS and SANs [68]. They discussed the localization characteristics and orientation of monomeric bacteriorhodopsin, a protein unit found in the purple membrane of multiple halobacteria, in the lipid nanodisc constituting POPC and 1-palmitoyl-2-oleoyl-sn-glycero-3-phospho-(1'-rac-glycerol) (POPG). It was observed that the protein orients itself in a slightly tilted fashion in the bilayer, and the thickness of the bilayer effectively reduces upon the addition of bacteriorhodopsin.

### 4.3 Biosensing Applications

Developing novel biosensors is an immensely crucial and tedious process requiring sensor materials to be extremely sensitive with a dynamic range. Lipid bilayers are an excellent option for sensing multiple biologically active molecules, including proteins, peptides, toxins, hormones, and enzymes. These bilayers are exploited in the form of SLBs, TLBs, and another class of biomimetic membranes known as black lipid membranes (BLMs), which are free-standing lipid planar bilayers. BLMs are generally formed across a small hole to study permeability across two different solutions and appear black upon reflection of light. Since their inception, these BLMs have been used to sense pore-forming biomolecules forming an ion channel across them. For example, in 1993, Dimitrios & Nikolelis formed black lipid membranes of egg phosphatidylcholine (PC) and dipalmitoylphosphatidic acid (DPPA) separating two compartments containing electrolytes and the corresponding ion currents were measured using electrodes [69]. This was followed by adding two different insecticides, monocrotophos and carbofuran, to one of the compartments. The generated ion channel current across the compartments was measured depending on the sensing activity of the BLM. In another study, Dimitrios and Vangelis used the same BLM setup to sense their interaction with atrazine, a chemical herbicide used to kill weeds [70]. It was observed that the intensity of the transient current signal was directly correlated to the concentration of atrazine. Although BLMs are rigorously explored for their sensing activities, these BLMs lack mechanical and electrical stability and, hence, often require some solid support and chemical modifications to overcome these barriers

SLBs or TLBs, with their well-defined mechanical supports and stability in aqueous environments, tend to be outstanding biosensors that offer acoustic, optical, and electrical detection of biological entities. Briand and group, combined quartz crystal microbalance with dissipation monitoring technique (QCM-D) and electrochemical impedance spectroscopy (EIS) to develop a novel biosensor [71]. This setup utilized SLBs coated on silicon dioxide surfaces for sensing and monitoring the interaction of a pore-forming peptide called gramicidin D. It was observed that increasing the peptide concentrations altered the membrane properties by causing structural changes as confirmed by the AFM measurements. In another study, Liu et al. improvised a device with SLB formation on conducting polymer electrodes using a solvent-assisted method [72]. This device contained a ligand for protein recognition on its SLB surface deposited on the electrode. Furthermore, fluorescence microscopy and electrochemical impedance spectroscopy verified the device's high specificity for distinct proteins. Su and the group followed a similar approach for forming mammalian and bacterial membranes mimicking SLBs on conducting polymer electrodes [73]. They quantified the interaction of these SLBs with a bacterial toxin (α-hemolysin) and an antibiotic compound (polymyxin B). The EIS modelling confirmed that α-hemolysin could form pores in the bilayer while polymyxin B disrupted the bilayer structure.

## 5. Conclusions and future prospects

In the last few decades, biomimetic membrane systems have been at the forefront of tackling multiple biological problems. These systems have gone from highly small-scaled, secluded, and complicated to broad, global, and simplified models for understanding numerous biological phenomena. In addition, these model systems have come together to form a strong base for designing advanced biosensing, bioelectronic, and bio-microelectromechanical (bio-MEMS) devices to pursue improved therapeutic and diagnostic applications. In this regard, through this chapter, we have demonstrated the conventional synthesis processes, limitations, and applications of multiple biomimetic models that have been traditionally used for progressing biomedical research.

Lipid monolayers, the most simplified form of biomimetic model membrane systems, actively contribute to exploring thermodynamical interactions and mechanism studies. Supported lipid bilayers and tethered lipid bilayers find multiple applications in sensing and membrane protein interactions. Further, liposomes in the form of unilamellar vesicles, multilamellar vesicles, and novel lipid nanodiscs have been regularly capitalized on for active molecule deliveries.

While these traditional methods for developing biomimetic systems have their distinct importance, the future holds innovative outlooks on improving their efficiency. For instance, in recent years, microfluidics-based synthesis techniques have been explored for developing liposomes. Additionally, the integration of nanotechnology and conventional membrane models has led to the development of lipid nanoparticles [74]. These lipid nanoparticles have also been designed in many forms, including solid lipid nanoparticles, nanostructured lipid carriers, lipid–drug conjugates, and lipid nanocapsules, offering effortless active drug delivery, making them suitable for pharmacological applications. Moreover, hybrid membrane-coated biomimetic nanoparticles have gained immense attention due to their immune evasion characteristics and targeting capability [75].

## Acknowledgments

AS thanks Council of Scientific & Industrial Research (Government of India) for the Senior Research Fellowship [CSIR-SRF Direct 09/1128(18228)/2024-EMR-I] and Shiv Nadar Institution of Eminence for the research fellowship. VKS and SKG thank Board of Research in Nuclear Sciences (BRNS), Department of Atomic Energy (DAE), Govt. of India for the financial support to conduct the research through a project with sanction number 58/14/13/2022-BRNS/37059.

**Figures**

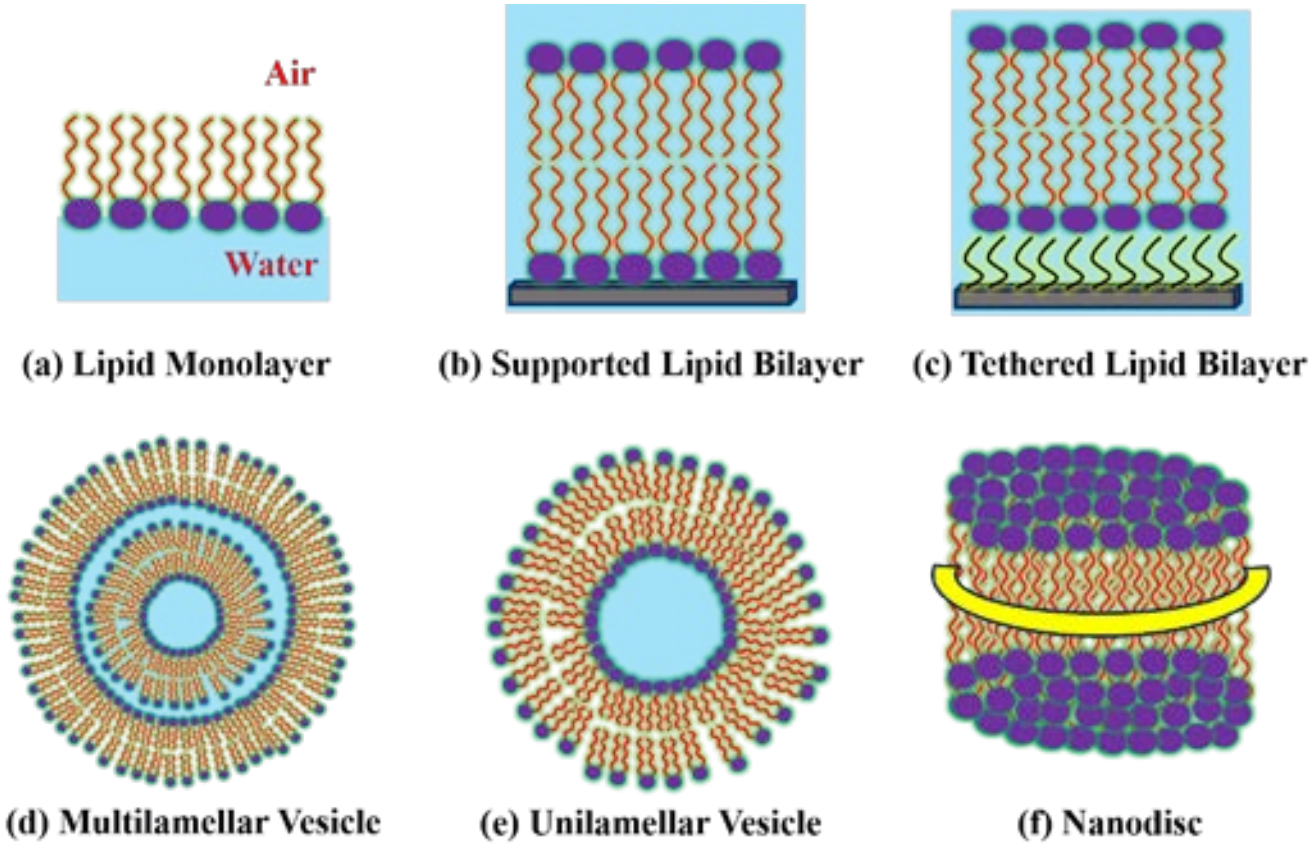

***Figure 1.*** *Common model biomembrane systems including (a) Lipid Monolayer, (b) Supported Lipid Bilayer (SLB), (c) Tethered Lipid Bilayer (TLB), (d) Unilamellar Vesicles, (e) Multilamellar Vesicles and, (f) Nanodisc.*

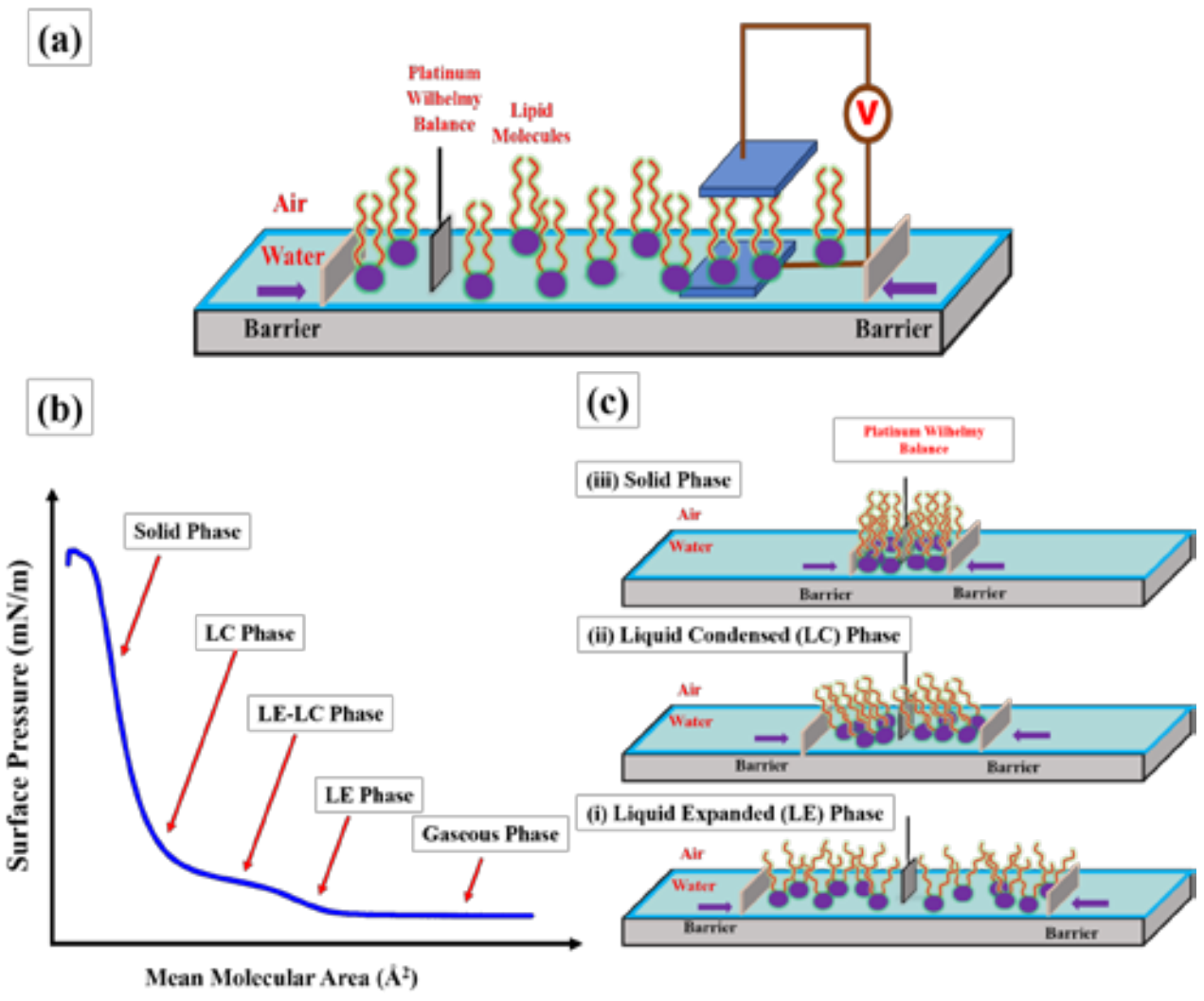

***Figure 2.*** *(a) Schematic of a Langmuir Blodgett (LB) trough measuring surface pressure and surface potential of a lipid monolayer with respect to the trough area;(b) A surface pressure-area isotherm of lipid molecules collected from the LB trough, displaying all the different phases of the lipid molecules during their self-assembly to form a monolayer; and (c) Molecular arrangement of the lipid molecules during the compression process in (i) liquid expanded (LE) phase, (ii) liquid condensed (LC) phase, and (iii) solid phase.*

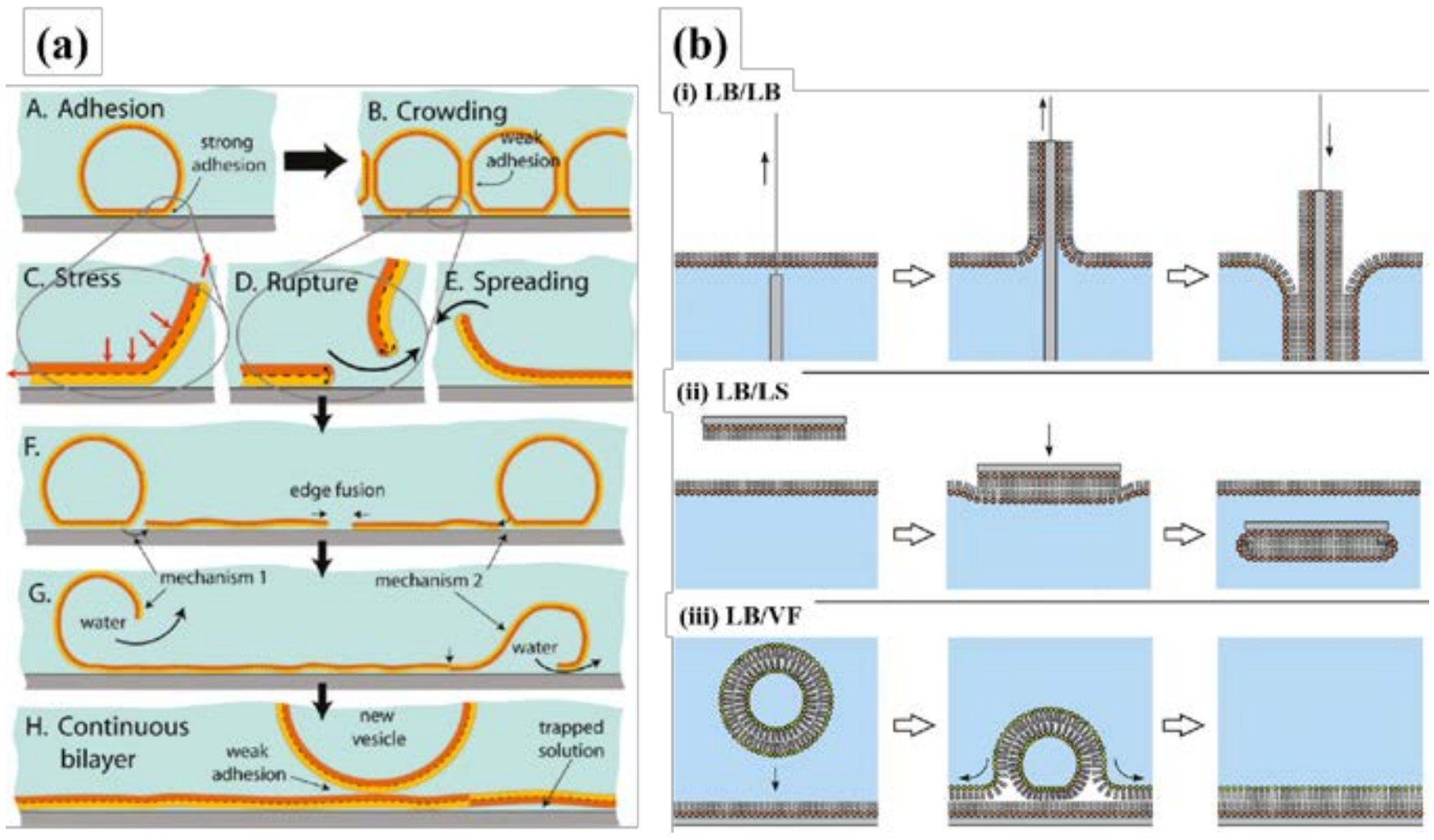

***Figure 3.*** *(a) Different stages of vesicle adsorption: (A) adhesion, (B) crowding, (CfE) stress-induced rupture and spreading of bilayer patches that can expose either leaflet by either mechanism 1 or 2, (F, G) coalescence of high energy edges and expulsion of water and excess lipid, and (H) growth of patches into a continuous bilayer; further adsorption of vesicles to the bilayer is weak and does not lead to their rupture or spreading. Figure 3(a) has been adapted with permission from reference [11] Copyright 2009, American Chemical Society; and (b) (i) Schematic LB deposition of solid-supported bilayers. (Left) After the substrate has been immersed, lipid is deposited on the air−water interface and compressed to the desired surface pressure. (Center) The substrate is drawn out of the subphase perpendicularly to the monolayer at the air−liquid interface to deposit the inner leaflet. (Right) To deposit the outer leaflet layer, the substrate is then lowered through the interface ;(ii) Langmuir−Schaefer (LS) method to horizontally deposit the outer leaflet onto a substrate-supported inner monolayer leaflet, (iii) Vesicle fusion (VF) on an LB-deposited monolayer. Figure 3(b) has been adapted with permission from reference [16] Copyright 2018, American Chemical Society.*

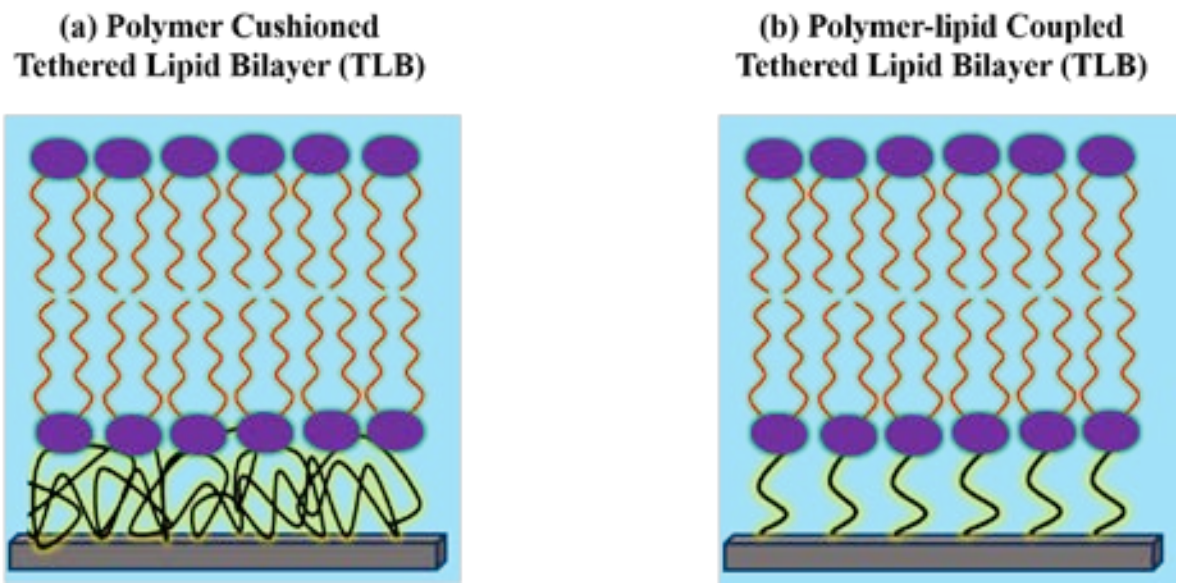

***Figure 4.*** *Schematic of (a) Polymer cushioned tethered lipid bilayer (TLB); and (b) Polymer-lipid coupled TLB.*

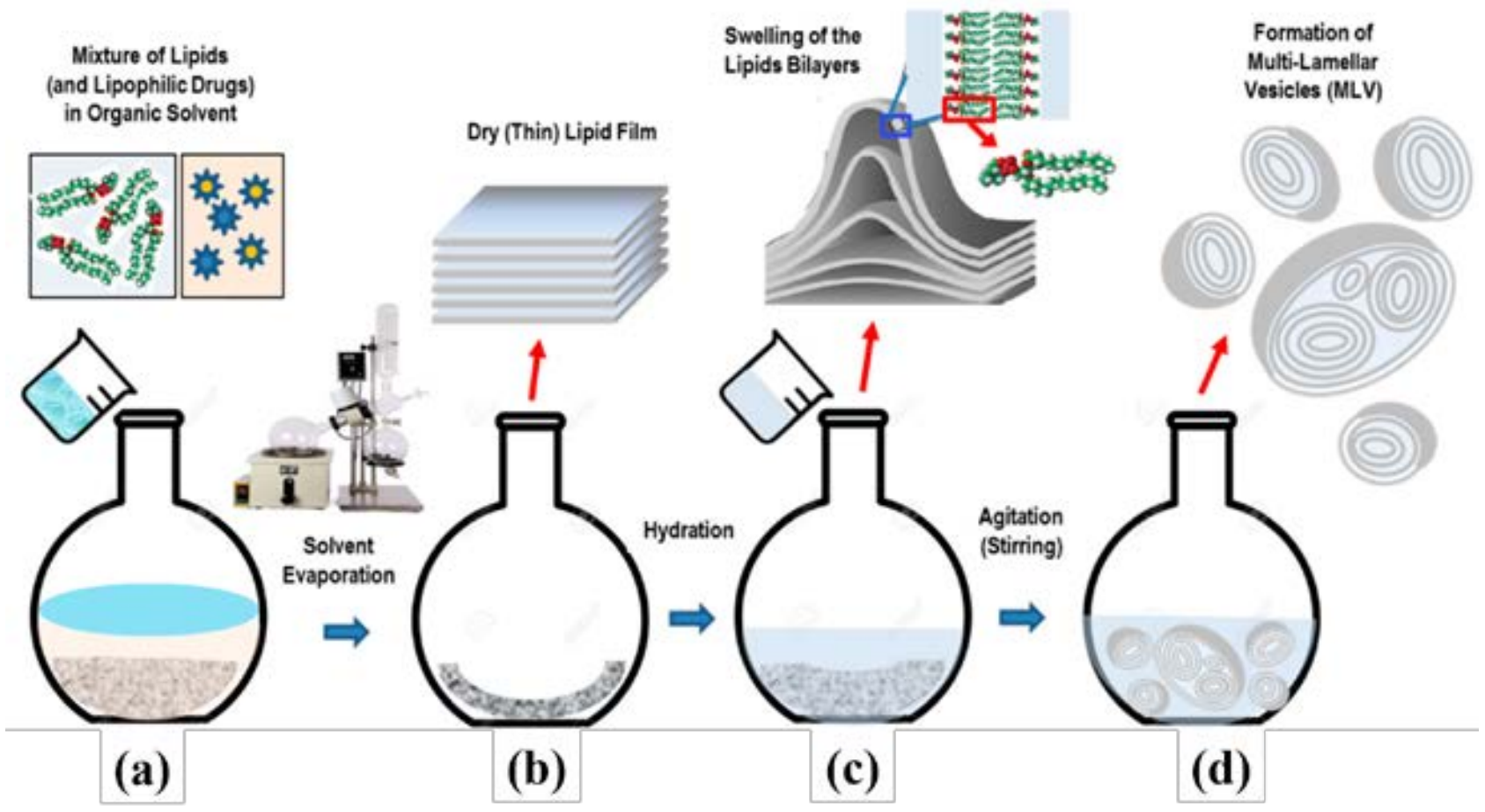

***Figure 5.*** *Schematic representation of the main stages of the thin-film hydration method of liposome preparation. The main lipid components (and eventually lipophilic drugs/macromolecules) are dissolved in organic solvent (a). After the evaporation of the solvent, a dry (thin) lipid film is formed (b). The lipid film is then rehydrated in a saline buffer (eventually containing hydrophilic dugs to be entrapped), causing a swelling of the lipid bilayers' stacks (c). The successive agitation/stirring of the sample favours the formation of (polydispersed) multilamellar vesicles (d). Reproduced with permission from reference [31].*

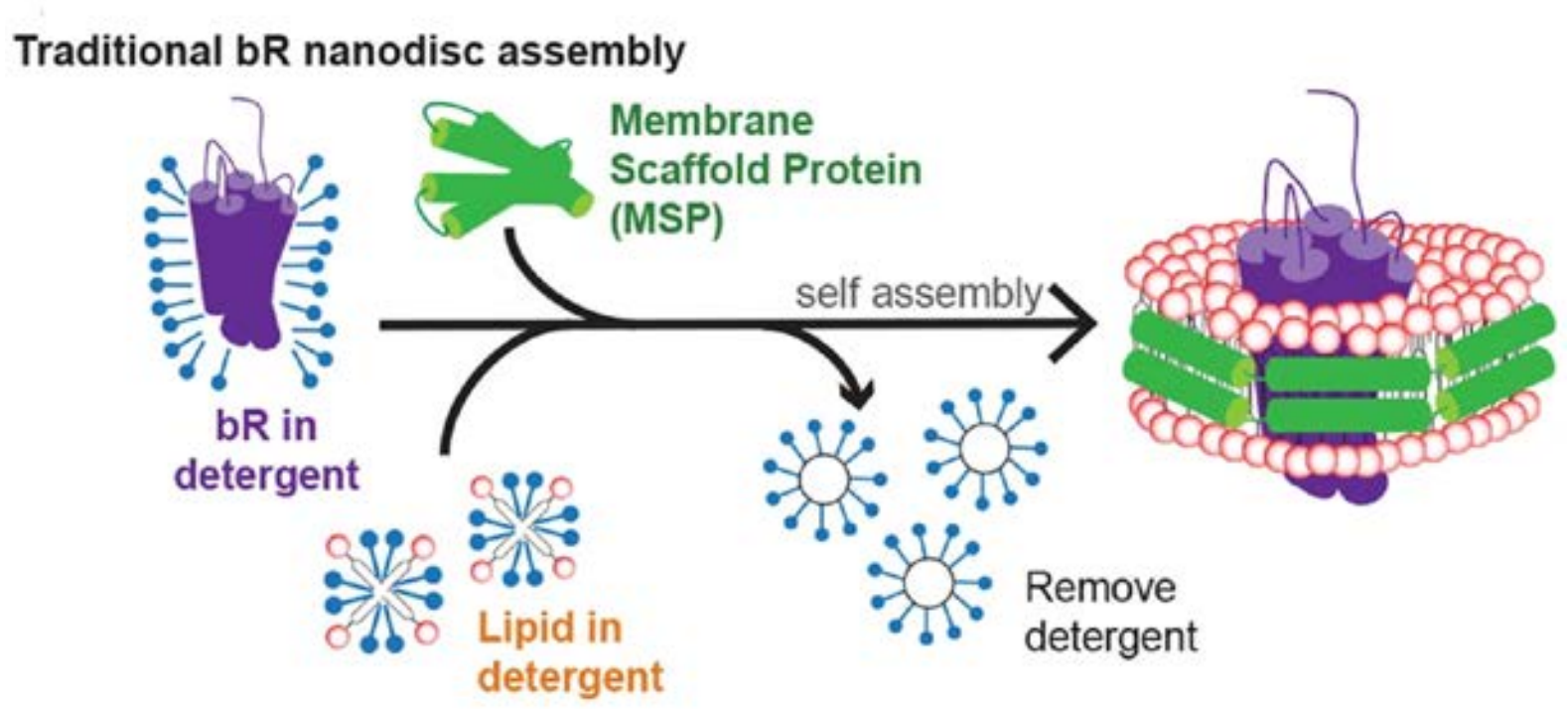

***Figure 6.*** *Schematic showing the formation of a phospholipid nanodisc enclosing the membrane protein Bacteriorhodopsin (bR). Reproduced with permission from reference [40].*

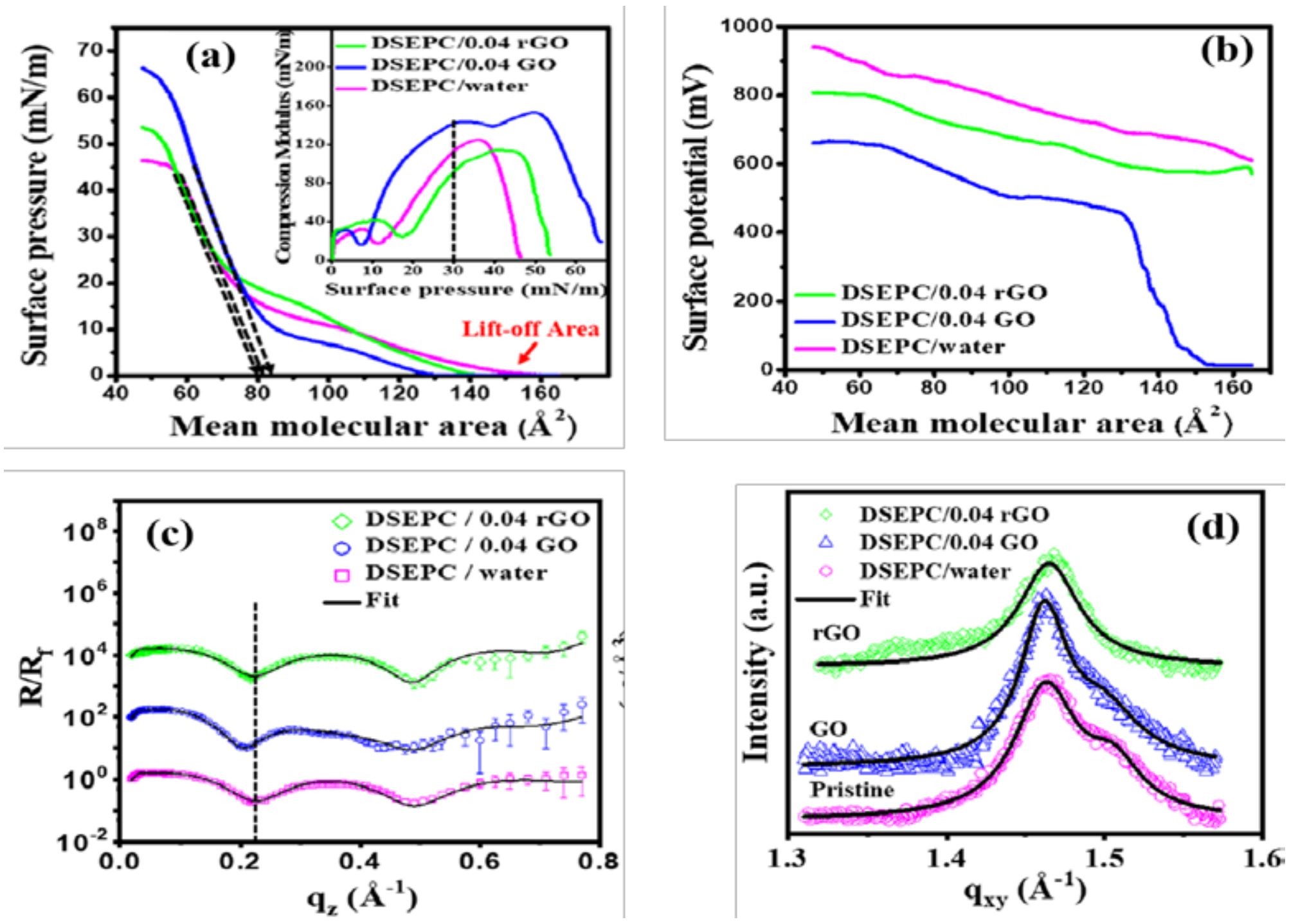

***Figure 7.*** *Interaction of a cationic lipid DSEPC monolayer with graphene oxide (GO) and reduced graphene oxide (rGO) quantified through (a) Surface pressure-area isotherms, (b) Surface potential-area isotherms, (c) X-ray reflectivity and, (d) Grazing incidence X-ray diffraction. Reproduced from reference [44] with permission from Royal Society of Chemistry.*

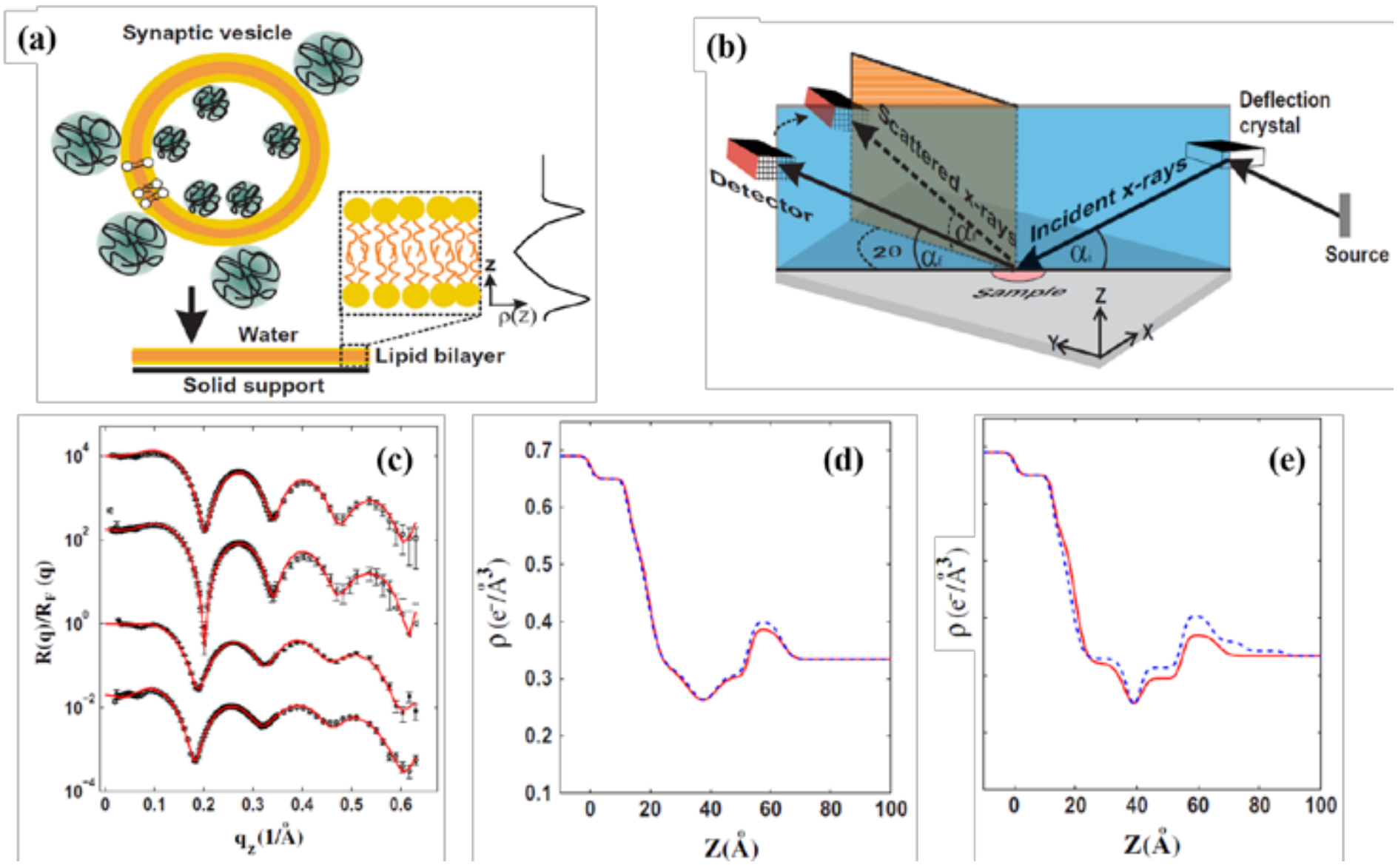

***Figure 8.*** *(a) Synaptic vesicles introduced on top of a supported lipid bilayer with the ρ(z) showing the electron density profile where z represents the z-axis is parallel to the sample normal; (b)Schematic illustration of the X-ray reflectivity measurements performed on the SLBs; (c) Fresnel normalized X-ray reflectivity curves for synaptic vesicles adsorbed on SLBs of POPC, with and without* $PIP_2$*; (d) Electron density profiles for a DOPC bilayer (solid line) and a DOPC bilayer with added synaptic vesicles (dashed line); and (e) Electron density profiles for a DOPC/PIP2 (4 : 1) bilayer (solid line) and for a DOPC/PIP2 (4 : 1)bilayer with added synaptic vesicles (dashed line). © Deutsche Physikalische Gesellschaft. Reproduced by permission of IOP Publishing reference [46].CC BY-NC-SA.*

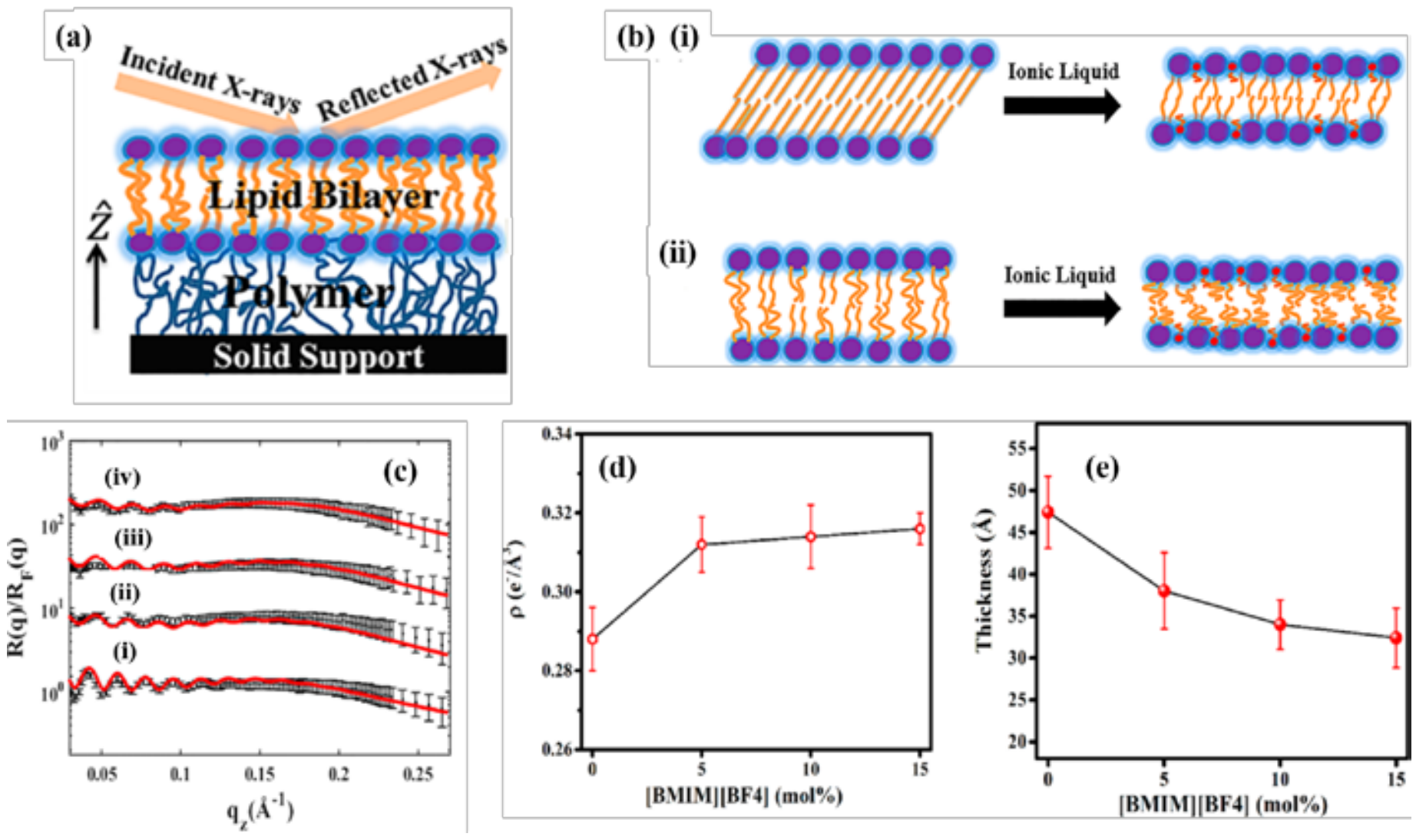

***Figure 9.*** *(a) Schematic of the tethered lipid bilayer ; (b) Effect of addition of an ionic liquid on the DPPC TLB in (i) Gel phase (35°C) and (ii) Fluid phase (48°C); (c) Fresnel normalized X-ray reflectivity data collected from DPPC SLBs at 48°C for (i) DPPC bilayer, DPPC bilayer with (ii) 5 mol%,(iii) 10 mol% and (iv) 15 mol% ionic liquid; (d) Electron density and (e) thickness of the DPPC lipid bilayer in the fluid phase (T = 48 °C) as a function of added [BMIM][BF4]. As the bilayer thickness decreases, the electron density of the layer increases. Adapted with permission from reference [47] Copyright 2017, American Chemical Society.*

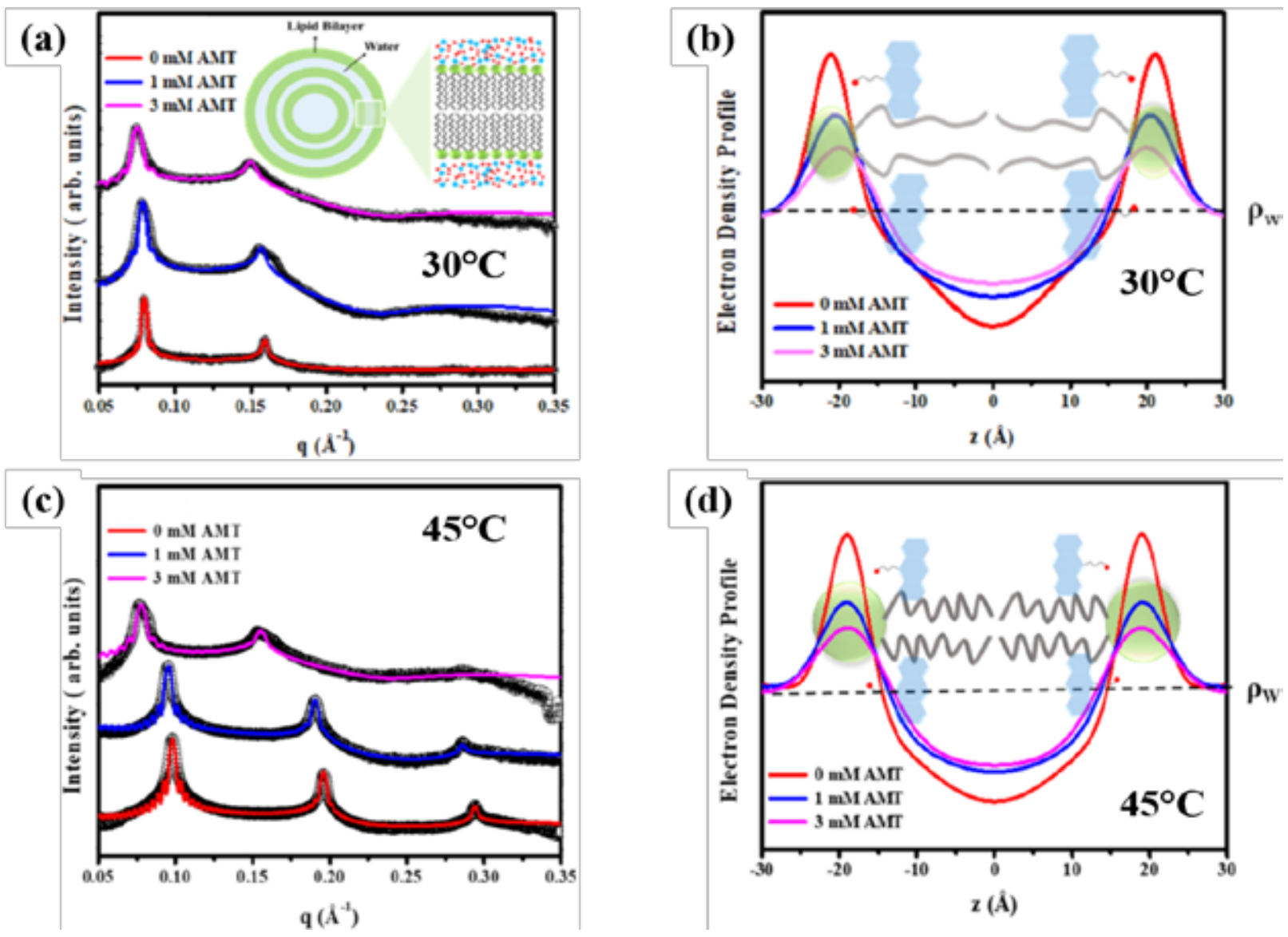

***Figure 10.*** *(a) Small angle X-ray scattering from multilamellar vesicles of brain sphingomyelin (BSM 18:0) in the presence of varied concentration of AMT at gel phase (30°C) and (c) fluid phase (45°C). The solid lines represent the model fits; and, the electron density profile originating from the fits of the data for (b) gel phase (30°C) and (d) fluid phase (45°C). Adapted with permission from reference [48].*

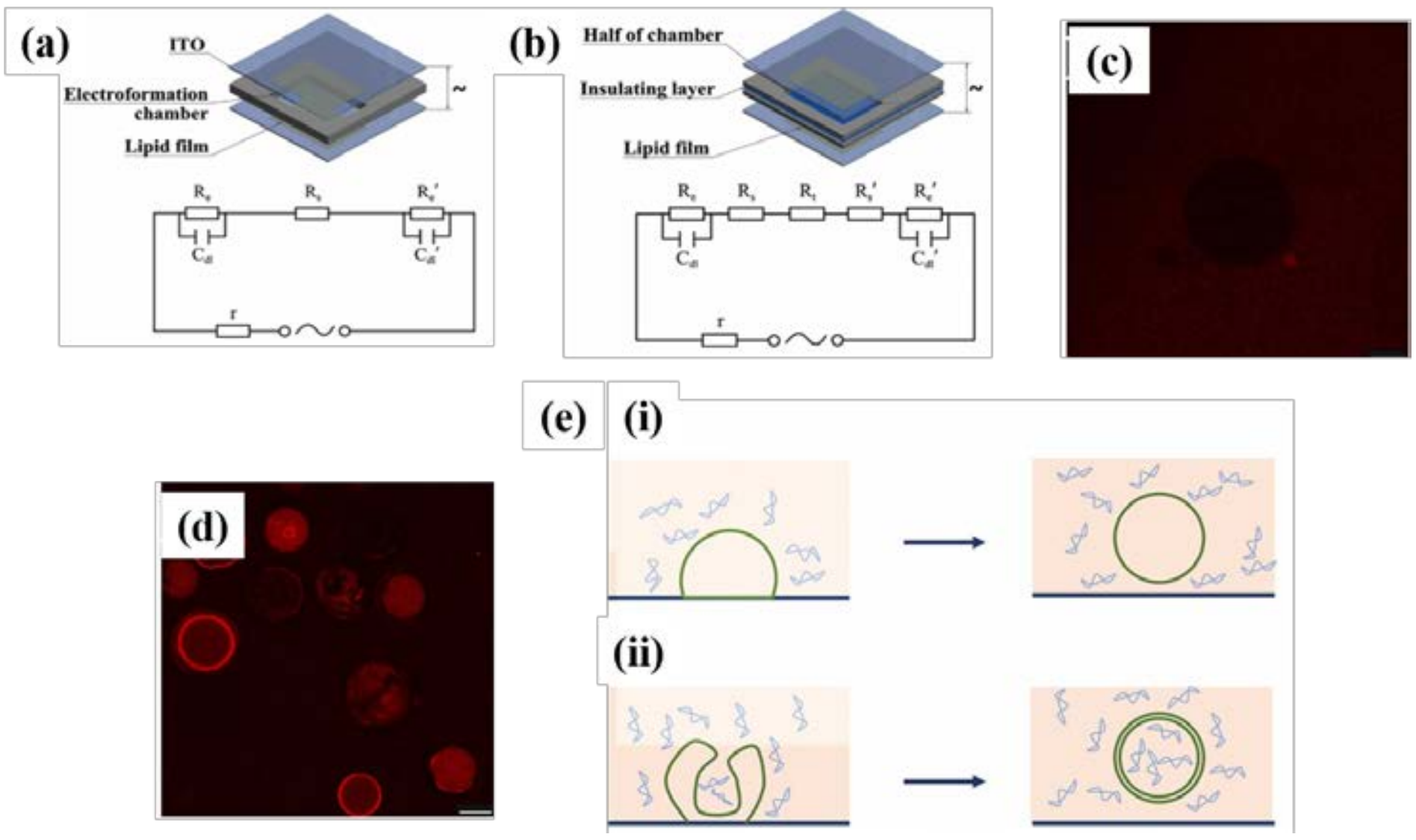

***Figure 11.*** *Improvement of GUV electroformation by the modified device. (a) Schematic diagram of the conventional electroformation device and equivalent circuit. (b) Schematic diagram of the modified electroformation device and equivalent circuit; an insulating layer was inserted in the middle of the chamber to separate the solution between the two electrodes; Laser confocal image of (c) vesicles prepared by the conventional device in plasmid DNA solution and (d) vesicles prepared by the modified device in plasmid DNA solution.; and (e) Schematic of vesicle formation in conventional device. (H) Schematic of vesicle formation and DNA encapsulation in the modified device. Adapted with permission from reference [54].*

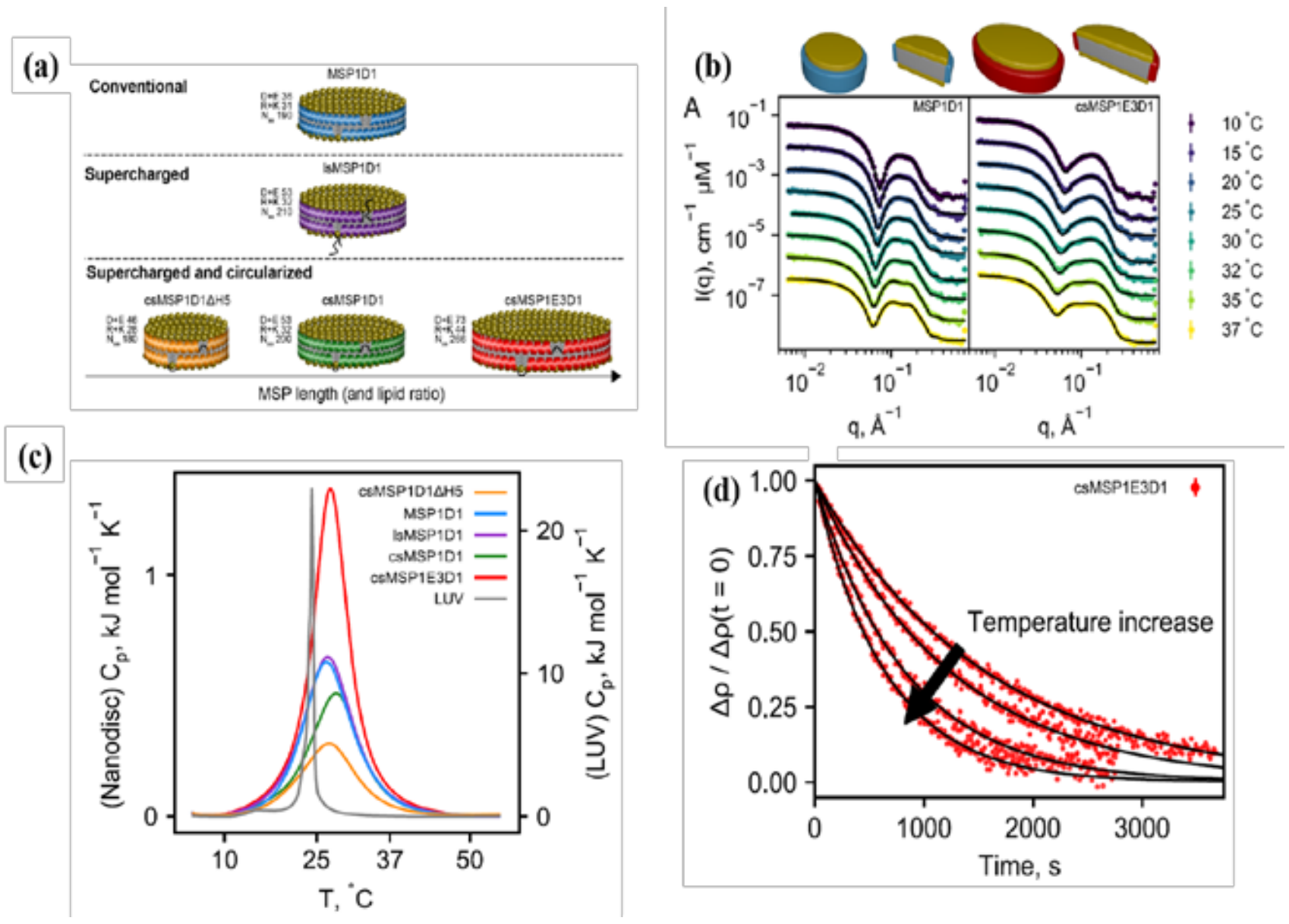

***Figure 12.*** *(a) Schematic displaying all the nanodiscs used with varying in size, charge, and circularization; (b) Small angle X-ray scattering data collected from 10 °C to 37 °C forMSP1D1-DMPC and csMSP1E3D1-DMPC for observing the temperature induced structural changes; (c) Differential scanning calorimetry thermogram for MSP-DMPC nanodiscs (left axis) and pure DMPC LUVs (right axis); and (d) Time resolved – small angle neutron scattering data collected on supercharged nanodisc at approximately 30 °C, 32 °C, 35 °C, and 37 °C, for observing the  dependence of lipid exchange dynamics on the size and charge of the nanodisc The SANS data were fitted by a single-exponential decay function (solid line) and subsequently normalized for visualization. Adapted with permission from reference [55]. Copyright 2021, American Chemical Society.*